# Meson Transition Form Factors From A QCD Model Field Theory [†]

P. C. Tandy

Center for Nuclear Research, Department of Physics, Kent State University,
Kent, Ohio 44242, USA

**Abstract**

We discuss form factors and coupling constants for the $\gamma^*\pi^0\gamma$, $\rho\pi\pi$ and $\gamma\pi\rho$ interactions generated by a model field theory that produces finite size $\bar{q}q$ meson modes. The approach implements dressing of the vertices and propagators consistent with dynamical chiral symmetry breaking, gauge invariance, quark confinement and perturbative QCD.

## 1 Introduction

We generate mesons as $\bar{q}q$ composites with intrinsic size from the Global Color Model (GCM). [1, 2] This is a phenomenologically successful QCD-based model field theory that permits an accessible and covariant description of meson substructure and interactions in terms of dynamically dressed quark degrees of freedom. The GCM action is

$$S[\bar{q},q] = \int d^4x\, \bar{q}(x)(\gamma\cdot\partial_x + m)q(x) + \frac{1}{2}\int d^4x d^4y\, j_\mu^a(x) D_{\mu\nu}(x-y) j_\nu^a(y) \qquad (1)$$

where $j_\mu^a(x) = \bar{q}(x)\frac{\lambda^a}{2}\gamma_\mu q(x)$. The formulation is in Euclidean space with metric $\delta_{\mu\nu}$, and in this work, we consider only two flavors ($u$ and $d$). The GCM represents the gluon sector by a finite range effective gluon two-point function ($D_{\mu\nu}(x) = \delta_{\mu\nu} D(x)$, in the Feynman-like gauge chosen here), thus formalizing the Abelian approximation to QCD. Nevertheless, the chiral anomalies [2] are properly embedded and this is important for several of the topics addressed here. A closely related model is that of Nambu and Jona-Lasinio [3] where the interaction is a contact one. In practical terms the difference is important. Quark confinement can be incorporated within the GCM and will prevent a spurious $\bar{q}q$ width for the heavier mesons. Finite size effects associated with dynamical quark self-energies and meson Bethe-Salpeter (BS) amplitudes provide natural regularization of many of the loop integrals for key physical quantities.

Studies of mesons and baryons within the GCM require hadronization techniques [4] and a recent review [5] is available. For mesons, a brief summary is the following. A functional change of path integration variables allows the second term of (1) to be eliminated in favor of $\bar{q}(x)\Lambda^\alpha \mathcal{B}^\alpha(x,y)q(y)$ and a term quadratic in $\mathcal{B}$. These auxiliary boson fields $\mathcal{B}^\alpha(x,y)$ transform as $\bar{q}(y)\Lambda^\alpha q(x)$ where $\Lambda^\alpha$ are the matrices from Fierz reordering of the current-current term. The saddle point configuration of $\mathcal{B}^\alpha$

---

yields a quark self-energy (equivalent to the ladder Dyson-Schwinger equation (DSE)) and expansion about there identifies a meson action in terms of field variables $\hat{\mathcal{B}}$ for propagating modes. The resulting bare inverse propagator may be used to generate eigenvectors $\bar{\Gamma}(q,P)$ for expansion of the bilocal fields as $\hat{\mathcal{B}}(q,P) \to \bar{\Gamma}_b(q,P)b(P)$, leaving an effective local field variable $b(P)$ for each meson mode. On the mass-shell, the $\bar{\Gamma}$ so defined are solutions of the ladder Bethe-Salpeter equation. The momentum $q$ is conjugate to $x - y$ and $P$ is conjugate to $\frac{x+y}{2}$.

To allow for electromagnetic (EM) coupling to the meson modes, the bosonization method can be generalized [6] to account for a background electromagnetic field minimally coupled ($\partial_\nu \to \partial_\nu - i\hat{Q}A_\nu$) to the bare quarks. The saddle point configuration $\mathcal{B}_0[A_\nu]$ now dresses the photon-quark vertex as well as the quarks. Gauge invariance of the EM coupling at the quark level can be translated to the meson level. [6] The method may be adapted for ease of computation, and for our purposes here, the following simplified version of the resulting meson action shall suffice

$$\hat{\mathcal{S}}[A,\pi,\rho,...] = \mathrm{Tr} \sum_{n=1}^\infty \frac{(-)^n}{n}[\tilde{S}(i\gamma_5\vec{\tau}\cdot\vec{\pi}\bar{\Gamma}_\pi + i\gamma_\mu\vec{\tau}\cdot\vec{\rho}_\mu\bar{\Gamma}_\rho + \cdots)]^n$$
$$+ \frac{9}{2}\int[\frac{1}{2}\pi\bar{\Gamma}_\pi D^{-1}\bar{\Gamma}_\pi\pi + \rho\bar{\Gamma}_\rho D^{-1}\bar{\Gamma}_\rho\rho + \cdots] + \delta\mathcal{S}[A,\rho,\omega]. \tag{2}$$

The trace here is over spin, flavor and color as well as space-time coordinates. We have employed only the dominant covariant (i.e. single Dirac matrix) for each meson mode. The EM field $A_\nu$ appears only in the dressed quark propagator $\tilde{S}[A_\nu]$ and in the term $\delta\mathcal{S}[A,\rho,\omega]$ which is linear in the indicated meson fields and is at most linear in $A_\nu$ in the more useful approaches. When $A_\nu = 0$, the $n = 1$ contribution from the first term cancels the $\delta\mathcal{S}$ term since in that case the meson fields are defined as fluctuations about the saddle point configurations. Then $S = \tilde{S}[0]$ has ladder DSE content giving $S(p)^{-1} = i\gamma\cdot p + \Sigma(p) + m$. When $A_\nu \neq 0$, rather than expand about the new saddle point configurations, we choose to expand about the configurations that produce $\tilde{S}(p,k)^{-1} = \delta(p-k)S(p)^{-1} + \Gamma_\nu(\frac{p+k}{2};p-k)A_\nu(p-k)$ where $\Gamma_\nu(p;q)$ is a dressed quark-photon vertex that is convenient for calculations. The third term of Eq. (2) depends on the choice of $\Gamma_\nu$ but it does not contribute to the interactions we study here.

A central element of the resulting effective action for bare hadrons is the dressed quark propagator. The necessary calculation of meson propagators and interaction vertex functions is complicated by the fact that the dressed quark propagator must be continued to complex Euclidean momenta to access the meson mass-shells in the timelike domain. In order to explore the capabilities and limitations of the overall approach, recent efforts have chosen the dressed quark propagator as the vehicle that carries the phenomenological input. Within the representation $S(p) = -i\gamma\cdot p\,\sigma_V(p^2) + \sigma_S(p^2)$, we employ the parametrized amplitudes [7]

$$\bar{\sigma}_S(x) = c\,e^{-2x} + \frac{1-e^{-b_1 x}}{b_1 x}\frac{1-e^{-b_3 x}}{b_3 x}\left(b_0 + b_2\frac{1-e^{-b_4 x}}{b_4 x}\right) + \frac{\bar{m}}{x+\bar{m}^2}(1-e^{-2(x+\bar{m}^2)}) \tag{3}$$

$$\bar{\sigma}_V(x) = \frac{2(x+\bar{m}^2) - 1 + e^{-2(x+\bar{m}^2)}}{2(x+\bar{m}^2)^2} - c\,\bar{m}\,e^{-2x}, \tag{4}$$

with $x = p^2/\lambda^2$, $\bar{\sigma}_S = \lambda\sigma_S$, $\bar{\sigma}_V = \lambda^2\sigma_V$, $\bar{m} = m/\lambda$, where $m$ is the bare quark mass and $\lambda$ is the momentum scale. This $S(p)$ is an entire function in the complex momentum plane, a condition sufficient to ensure the absence of quark production thresholds in S-matrix amplitudes for physical processes. [8] This form is guided by a confining model DSE [9] and by the behavior found in realistic DSE studies. [10] It is consistent with pQCD in the deep Euclidean region up to $\ln(p^2)$ corrections. The parameters are $\lambda = 0.568$ GeV, $c = 0.0406$, $m = 6.7$ MeV, and $(b_0,b_1,b_2,b_3,b_4) = (0.118, 2.51, 0.525, 0.169, 1\times 10^{-4})$. The fitted soft chiral physics quantities produced by this parameterization are [7] $f_\pi = 92.4$ MeV, $<\bar{q}q>_{1\text{GeV}^2} = (232 \text{ MeV})^3$, $m_\pi = 140$ MeV, $r_\pi^{em} = 0.54$ fm together with reproduction of the experimental $\pi\pi$ scattering lengths to within 20%.

The choice of quark propagator parameters is equivalent to an implicit choice of effective gluon propagator and vertex in the underlying DSE dynamics. [10] Ghost contributions in DSE studies have been studied in Landau gauge and shown not to modify the qualitative features of quark and gluon propagators.[11] Quantitatively, ghosts provide a small ($< 10\%$) effect. We therefore expect that a well-chosen phenomenological quark propagator captures those aspects of the underlying dynamics that is of practical importance for the chiral physics quantities that guide it. Independent applications, such as those contained here, provide important additional tests of this standpoint.

In terms of the equivalent representation $S(p)^{-1} = i\gamma \cdot p A(p^2) + B(p^2) + m$, the chiral limit for the pion mass-shell Bethe-Salpeter amplitude is $\Gamma_\pi(p; P^2 = 0) = i\gamma_5 \tau B(p^2, m=0)/f_\pi$ since the DSE for $B(p^2)$ and the Bethe-Salpeter equation for the pion invariant amplitude become identical. [12] Hence the pion is both a $\bar{q}q$ bound state and a Goldstone boson. For finite $m$ we use the pseudoscalar form $\Gamma_\pi(p; P) \approx i\gamma_5 \tau B(p^2, m)/f_\pi$ consistent with PCAC. We use the Ball-Chiu [13] Ansatz for the dressed quark photon vertex which is $\Gamma_\mu(p;q) = \hat{Q}\bar{\Gamma}_\mu(p;q)$, where $\hat{Q} = \frac{1}{2}(\tau_3 + \frac{1}{3})$ is the quark charge operator, and

$$\bar{\Gamma}_\mu(p;q) = -i\gamma_\mu \frac{1}{2}\Big(A(p_+) + A(p_-)\Big) + \frac{p_\mu}{p \cdot q}\Big[i\gamma \cdot p\Big(A(p_-) - A(p_+)\Big) + \Big(B(p_-) - B(p_+)\Big)\Big] \quad (5)$$

with $p_\pm = p \pm \frac{q}{2}$. This vertex satisfies the Ward-Takahashi identity, transforms correctly and has the correct perturbative limit; but it is not unique. We have investigated several other choices [10] and found no significant change in our conclusions.

## 2  The $\pi^0\gamma\gamma$ Form Factor

The pion charge form factor for space-like momenta is one of the simplest but non-trivial testing grounds for applications of QCD to hadronic properties. A closely related quantity that has received less attention is the $\gamma^*\pi^0 \to \gamma$ transition form factor. [14] Here the photon momentum dependence maps out a particular off-shell extension of the axial anomaly. [15] Presently available data for this transition form factor in the space-like region $Q^2 < 2.5$ GeV$^2$ is from the CELLO [16] collaboration at the PETRA storage ring where the process $e^+e^- \to e^+e^-\pi^0$ was measured with geometry requiring one of the two intermediate photons to be almost real. There is renewed interest in this transition form factor due to the prospect of obtaining higher precision data over a broader momentum range via virtual Compton scattering from a proton target at CEBAF. [17] Features distinguishing the present work [18] from a previous quark loop study [19] are: quark confinement (thus eliminating spurious quark production thresholds), dressing of the photon-quark vertex, and the dynamical relation between the pion Bethe-Salpeter amplitude and the quark propagator. The latter elements are crucial for producing the correct mass-shell axial anomaly independently of model details.

The relevant part of the action (2) is identified by expansion to second order in $A_\nu$. Only the first term (with $n = 1$) contributes and we find $\mathcal{S}[\pi^0\gamma\gamma] = -\text{Tr}[S\Gamma_\mu A_\mu S\Gamma_\nu A_\nu S\Gamma_\pi \pi^0]$. That is

$$\mathcal{S}[\pi^0\gamma\gamma] = \int \frac{d^4P d^4Q}{(2\pi)^8} A_\mu(-P-Q)A_\nu(Q)\pi^0(P)\Lambda_{\mu\nu}(P,Q), \quad (6)$$

where the vertex function is given by the integral

$$\Lambda_{\mu\nu}(P,Q) = -\text{tr}\int \frac{d^4k}{(2\pi)^4} \quad S(k-P-Q)\Gamma_\mu(k - \frac{P}{2} - \frac{Q}{2}; -P-Q)S(k)$$

$$\times \Gamma_\nu(k - \frac{Q}{2}; Q)S(k-Q)\Gamma_\pi(k - \frac{P}{2} - Q; P). \quad (7)$$

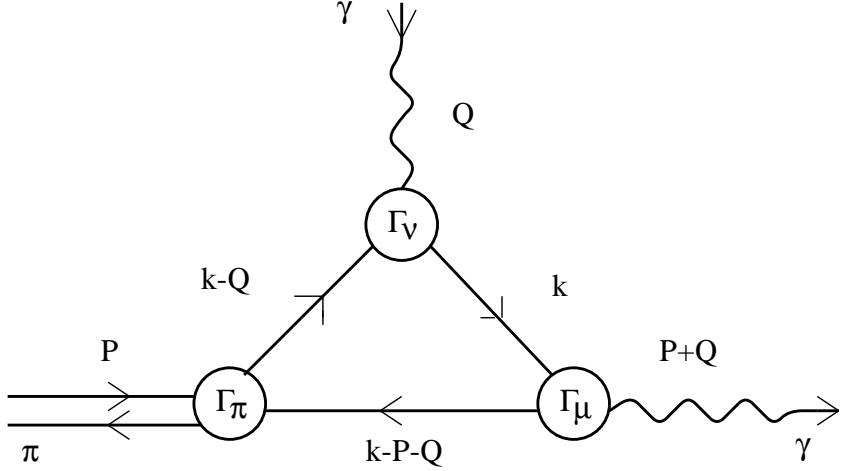

Figure 1: The quark triangle diagram for the generalized impulse approximation to the $\gamma^*\pi^0\gamma$ vertex.

The momentum assignments are shown in the quark triangle diagram of Fig. 1. The general form of the vertex allowed by CPT is $\Lambda_{\mu\nu}(P,Q) = -i\frac{\alpha}{\pi f_\pi}\epsilon_{\mu\nu\alpha\beta}P_\alpha Q_\beta\, g(Q^2, P^2, P\cdot Q)$ where $\epsilon_{4123} = 1$, $\alpha$ is the fine-structure constant, $f_\pi$ is the pion decay constant, and $g$ is the off-mass-shell invariant amplitude. With the one photon mass-shell condition $(P+Q)^2 = 0$, the invariant amplitude, denoted by $g(Q^2, P^2)$, is the object of the present work. For a physical pion the shape of the $\gamma^*\pi^0 \to \gamma$ transition form factor is given by $g(Q^2, -m_\pi^2)$. The chiral limit for the physical $\pi^0 \to \gamma\gamma$ decay amplitude is fixed at $\frac{\alpha}{\pi f_\pi}$ by the axial anomaly [20] which gives an excellent account of the 7.7 eV width and requires $g(0,0) = 1/2$. This follows only from gauge invariance and chiral symmetry in quantum field theory and provides a stringent check upon model calculations.

The $\pi^0\gamma\gamma$ vertex function in (7) is now completely specified in terms of the quark propagator. An exactly parallel situation holds for the spacelike pion charge form factor and this model has previously been shown to provide an excellent description of the data. [7] No adjustment of parameters is made in the present application. At $Q^2 = 0$, our numerical evaluation yields $g_{\pi^0\gamma\gamma} = g(0, -m_\pi^2) = 0.496$, in agreement with the previous application of this model [7] and in good agreement with the experimental value $0.504 \pm 0.019$. The chiral limit of this approach has been shown [7] to correctly incorporate the exact result $g(0,0) = 1/2$ produced by the axial anomaly *independent* of the form and details of the quark propagator.

The obtained form factor $F(Q^2) = g(Q^2, -m_\pi^2)/g(0, -m_\pi^2)$ at the pion mass-shell is displayed as $Q^2 F(Q^2)$ in Fig. 2 by the solid line along with the CELLO data. [16] We calculate a "radius" or interaction size, defined via $\langle r^2_{\gamma\pi^0\gamma}\rangle = -6\, F'(Q^2)|_{Q^2=0}$, of 0.47 fm while a monopole fit to the data yields [16] $0.65 \pm 0.03$ fm. The insert compares our low $Q^2$ result with a recent result from a QCD sum rule approach [21], and a monopole form [22] that interpolates from the leading asymptotic behavior $F(Q^2) \to 8\pi^2 f_\pi^2/Q^2$ argued from the pQCD factorization approach. [23] In the latter two approaches there is ambiguity due to: A) the unknown momentum scale at which perturbative behavior should set in; and B) assumptions for the pion wavefunction and how it should evolve with the momentum scale. [21] Within the present approach, both the photon coupling and the pion wavefunction evolve with $Q^2$ in a way determined by the evolution of the dressed quark propagators. This produces, in a single expression, both the ultra-violet behavior required by pQCD and the infra-red limit dictated by the axial anomaly. We note that the employed pion Bethe-Salpeter amplitude $B(p^2, m)/f_\pi$ has the correct leading power law behavior $m\lambda^2/p^2 f_\pi$ which implements the hard gluon contribution that dominates pQCD.

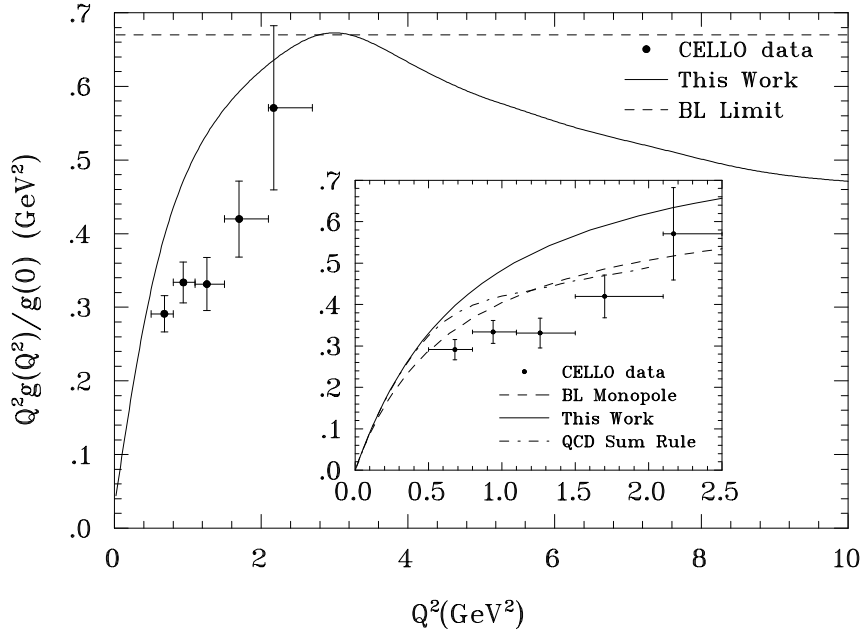

Figure 2: The $\gamma^*\pi^0\gamma$ transition form factor.

The dotted straight line in Fig. 2 is the pQCD factorization [23] limit $Q^2 F(Q^2) \to 8\pi^2 f_\pi^2 = 0.67$ GeV$^2$. Although our non-factorized calculation reaches this value near $Q^2 = 3$ GeV$^2$, we find a slow decrease with higher $Q^2$ consistent with a logarithmic correction. An excellent fit to the numerical results for 3.3 GeV$^2 \leq Q^2 \leq 10$ GeV$^2$ is provided by $F(Q^2) = A\,[1.0 + B\,Q^2\,\ln(C\,Q^2)]^{-1}$, where $A = 1.021$, $B = 0.461/m_\rho^2 = 0.777$ GeV$^{-2}$ and $C = 1.16/m_\rho^2 = 1.45$ GeV$^{-2}$. (We have neglected the anomalous dimension of the quark propagator in this calculation, which would modify the power of the ln-correction.)

The logarithmic correction to the anticipated $1/Q^2$ asymptotic behavior can be attributed to the persistent nonperturbative nature of the coupling to the final state soft photon in this exclusive process. [18] Numerically we find that, if a bare coupling were to be used for both photons as is implicit in the pQCD factorization approach, $F(Q^2)$ would eventually approach $8\pi^2 f_\pi^2/Q^2$. The turn-over in $Q^2 F(Q^2)$ near 3 GeV$^2$ predicted in Fig. 2 is barely within the $Q^2$ limit of 4 GeV$^2$ anticipated for measurements at CEBAF if a 6 GeV electron beam becomes available. [17] This turn-over and the logarithmic corrections generated by the loop integral are features also found in the parallel approach to the pion charge form factor.[7] For a study of the behavior of the vertex function off the pion mass shell, as needed for the anticipated CEBAF experiment, see Ref. [18].

## 3  The $\rho\pi\pi$ Form Factor

From (2) we identify $\mathcal{S}\,[\rho\pi\pi] = -\text{Tr}\,\left[Si\gamma_\mu\vec{\tau}\cdot\vec{\rho}_\mu\bar{\Gamma}_\rho(Si\gamma_5\vec{\tau}\cdot\vec{\pi}\bar{\Gamma}_\pi)^2\right]$, which yields

$$\mathcal{S}\,[\rho\pi\pi] = i\int\frac{d^4P,Q}{(2\pi)^8}\vec{\rho}_\mu(Q)\cdot\vec{\pi}(-P-\frac{Q}{2})\times\vec{\pi}(P-\frac{Q}{2})\Lambda_\mu(P,Q) \qquad (8)$$

where the vertex is

$$\Lambda_\mu(P,Q) = \int\frac{d^4k}{(2\pi)^4}\bar{\Gamma}_\rho(k+\frac{P}{2};Q)\bar{\Gamma}_\pi(k+\frac{Q}{4};-P-\frac{Q}{2})\bar{\Gamma}_\pi(k-\frac{Q}{4};P-\frac{Q}{2})T_\mu(k,P,Q) \qquad (9)$$

with

$$T_\mu(k,P,Q) = \text{tr}\left[S(k+\frac{P}{2}+\frac{Q}{2})i\gamma_\mu S(k+\frac{P}{2}-\frac{Q}{2})i\gamma_5 S(k-\frac{P}{2})i\gamma_5\right]. \tag{10}$$

From symmetry properties, it is not difficult to show that $\Lambda_\mu(P,Q) = -\Lambda_\mu(-P,Q) = \Lambda_\mu(P,-Q)$, which requires the general form

$$\Lambda_\mu(P,Q) = -P_\mu F_{\rho\pi\pi}(P^2, Q^2, (P\cdot Q)^2) - Q_\mu P\cdot Q H_{\rho\pi\pi}(P^2, Q^2, (P\cdot Q)^2). \tag{11}$$

With both pions on the mass-shell, $(P-\frac{Q}{2})^2 = (P+\frac{Q}{2})^2 = -m_\pi^2$. Equivalently, $P\cdot Q = 0$ and $P^2 = -m_\pi^2 - \frac{Q^2}{4}$ so that only the first term of (11) survives. The corresponding form factor $F_{\rho\pi\pi}(Q^2)$ contains the coupling constant as its mass-shell value, i.e. $F_{\rho\pi\pi}(Q^2 = -m_\rho^2) = g_{\rho\pi\pi}$. If the form factor is held at this value for all momenta, one obtains the point coupling limit which can be expressed in the standard form

$$\mathcal{S}[\rho\pi\pi] = -g_{\rho\pi\pi}\int d^4x\, \vec{\rho}_\mu(x)\cdot\vec{\pi}(x)\times\partial_\mu\vec{\pi}(x). \tag{12}$$

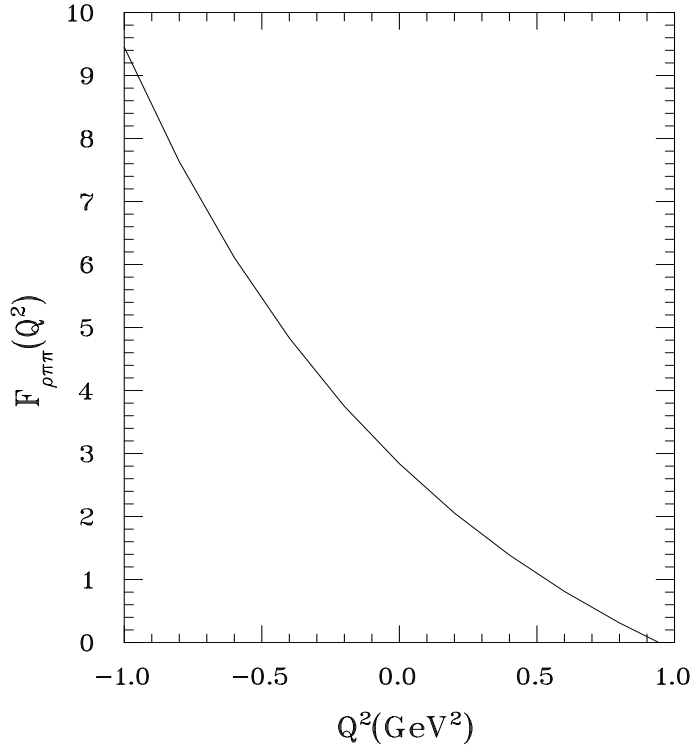

Figure 3: Behavior of the $\rho\pi\pi$ form factor with $\rho$ momentum in the viscinity of the mass-shell at $-0.6$ GeV$^2$.

The previous study [24] within the GCM made the approximation $g_{\rho\pi\pi} \approx F_{\rho\pi\pi}(P=Q=0)$ in order to avoid the occurrence of complex momenta in the arguments of the propagators and vertex functions in the integral (9). Complex momenta are unavoidable when a Euclidean formulation is continued to handle timelike external momenta. However when propagators and BS amplitudes are produced numerically it is a difficult and lengthy matter to extend such solutions to the complex plane. The utility of the parameterized representation, (3) and (4), of the quark propagator is that the continuations required to properly implement mass-shell constraints as needed for extraction of $g_{\rho\pi\pi}$ and other quantities are a simple matter. Since the propagator parameters have so far been constrained only by pion physics, there is no guarantee that continuation of $S(p)$ away from the real $p^2$ axis by an amount proportional to $m_\rho^2$ will be correct. The present exploratory calculations involving vector mesons constitute a first investigation of this matter.

The investigation here is limited since we have not made an independent calculation of the $\rho$ BS amplitude $\bar\Gamma_\rho(p)$. This requires knowledge of the effective gluon propagator $D(r)$ compatible, in a ladder DSE sense, with the given quark propagator. Although some progress has recently been made in this direction, [25, 26] in the present work we directly parameterize $\bar\Gamma_\rho(p^2) \propto e^{-p^2/a^2}$. This has proved to be effective in a related study. [27] The strength is set by the canonical normalization condition [20] which is equivalent to ensuring that the residue at the pole of the $\rho$ propagator is unity. Only the quadratic $\rho$ dependence from the first term of (2), i.e. the quark loop contribution to the $\rho$ inverse propagator, is needed to implement this. The range $a$ is then adjusted to reproduce the empirical value $g_{\rho\pi\pi}^{\rm expt} = 6.2$. This requires $a = 0.208$ GeV$^2$. The corresponding $\rho \to \pi\pi$ decay width, given by

$$\Gamma_{\rho\to\pi\pi} = \frac{g_{\rho\pi\pi}^2}{4\pi}\frac{m_\rho}{12}\left[1 - \frac{4m_\pi^2}{m_\rho^2}\right]^{3/2}, \tag{13}$$

is 151 MeV. The calculated form factor $F_{\rho\pi\pi}(Q^2)$ is shown in Fig. 3 for timelike and spacelike momenta in the viscinity of the mass-shell. The main conclusion from this calculation is that the previous approximation of using zero momentum to extract a coupling constant [24] can underestimate the value by almost a factor of 2.

## 4   The $\gamma\pi\rho$ Form Factor

The empirical $\rho$ BS amplitude set by $g_{\rho\pi\pi}$ as above enables a parameter-free prediction for the $\gamma\pi\rho$ vertex. Apart from being a consistency check in this manner, the $\gamma\pi\rho$ interaction together with the $\gamma\pi\pi$, $\gamma\gamma\pi$, and the $\rho\pi\pi$ processes, provide important guidance for extending the present approach to nonperturbative QCD modeling of meson physics to phenomena not dictated by chiral symmetry. Within nuclear physics, the associated isoscalar $\gamma^*\pi\rho$ meson-exchange current contributes significantly to electron scattering from light nuclei. In particular, our understanding of the deuteron EM structure functions for $Q^2 \approx 2 - 4$ GeV$^2$ is presently hindered by uncertainties in the behavior of this form factor. [28].

Expansion of (2) to first order in the EM field yields the $\gamma\pi\rho$ interaction as the pair of contributions

$$\mathcal{S}[\gamma\pi\rho] = -{\rm Tr}[S\Gamma_\mu A_\mu S i\gamma_5 \vec\tau \cdot \vec\pi \bar\Gamma_\pi S i\gamma_\nu \vec\tau \cdot \vec\rho_\nu \bar\Gamma_\rho] - {\rm Tr}[S\Gamma_\mu A_\mu S i\gamma_\nu \vec\tau \cdot \vec\rho_\nu \bar\Gamma_\rho S i\gamma_5 \vec\tau \cdot \vec\pi \bar\Gamma_\pi]. \tag{14}$$

With a vertex function $\Lambda_{\mu\nu}(P,Q)$ defined by

$$\mathcal{S}[\gamma\pi\rho] = -\int \frac{d^4P, Q}{(2\pi)^8} A_\nu(Q)\vec\pi(-P-\frac{Q}{2})\cdot\vec\rho_\mu(P-\frac{Q}{2})\Lambda_{\mu\nu}(P,Q), \tag{15}$$

one may combine the two terms in (14) to obtain the integral

$$\begin{aligned}\Lambda_{\mu\nu}(P,Q) &= \frac{e}{3}\int\frac{d^4k}{(2\pi)^4}\Gamma_\pi(k+\frac{Q}{4}; -P-\frac{Q}{2})\Gamma_\rho(k-\frac{Q}{4}; P-\frac{Q}{2})\\ &\times {\rm tr}[S(k_+ -\frac{Q}{2})\Gamma_\nu(k_+; Q)S(k_+ + \frac{Q}{2})i\gamma_5 S(k_-)i\gamma_\mu].\end{aligned} \tag{16}$$

Here $k_\pm = k \pm \frac{P}{2}$. The $\gamma\pi\rho$ EM current $(-(2\pi)^4 \frac{\delta\mathcal{S}}{\delta A_\nu(Q)})$ will be conserved as a consequence of maintaining EM gauge invariance at the quark level. Explicitly, use of the Ward-Takahashi identity $Q_\nu\Gamma_\nu(k;Q) =$ $\hat{Q}S^{-1}(k-\frac{Q}{2}) - \hat{Q}S^{-1}(k+\frac{Q}{2})$, in (16) immediately gives $Q_\nu\Lambda_{\mu\nu}(P,Q) = 0$. The general form of the vertex function can be shown by symmetries to be $\Lambda_{\mu\nu}(P,Q) = -i\frac{e}{m_\rho}\epsilon_{\mu\nu\alpha\beta}P_\alpha Q_\beta \; g_{\rho\pi\gamma} \; f(Q^2, P^2, P\cdot Q)$, as is expected for a coupling arising from the chiral anomaly. We have used the standard definition of the coupling constant so that, at the triple mass-shell point, the form factor $f = 1$.

Our numerical evaluation yields $g_{\gamma\pi\rho} = 0.5$. The experimental $\rho^+ \to \pi^+\gamma$ partial width ($67 \pm 7$ keV) determines the empirical value $g_{\gamma\pi\rho}^{\rm expt} = 0.54 \pm 0.03$. The $\gamma\pi\rho$ form factor obtained with on-mass-shell $\pi$ and $\rho$, and weighted by $Q^2$, is shown by the solid curve in Figure 4. Also shown for comparison is the vector dominance model (VDM) phenomenology initially used to include such meson exchange effects in electron scattering analysis, [29] as well as the result from a free constituent quark loop with bare photon coupling. [28, 30] The quark-based results produce a much softer form factor than is produced by the VDM assumption. Above 50 fm$^{-2}$, which is readily accessible in electron scattering, the differences are serious.

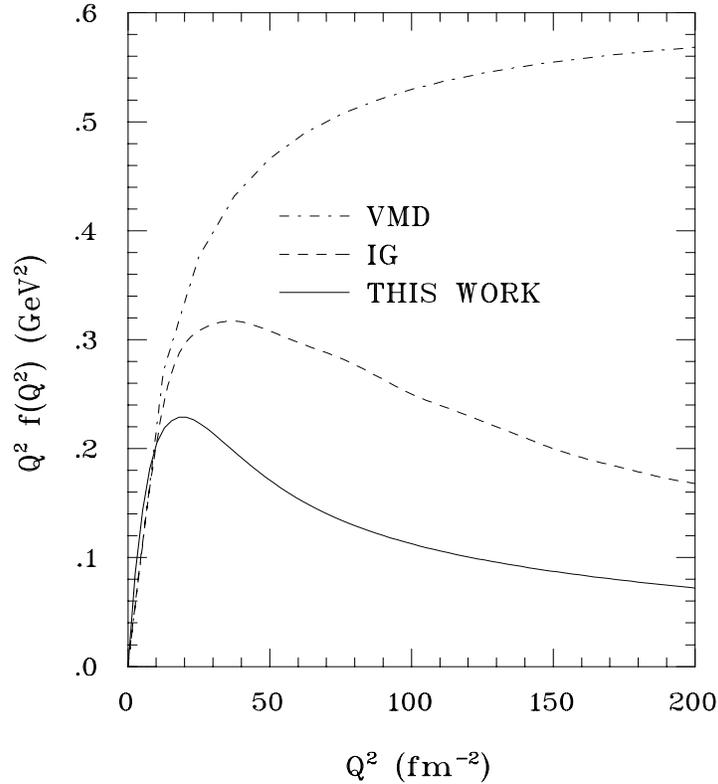

Figure 4: The photon momentum dependence of the $\gamma\pi\rho$ form factor from the on-shell approximation for both mesons.

## 5  Discussion

The photon-quark vertex employed in this work does not contain an explicit vector meson pole. The absence of the vector meson dominance mechanism in the explicit photon vertex is, in fact, consistent with the tree-level nature of the action (2). The $n = 1$ part of the first term, in combination with the third term, produces direct coupling of the photon with vector mesons. When the analysis is carried to first-order in meson loops, EM coupling mediated by a vector meson propagator will be produced consistently as an additional mechanism to that considered here. More detailed exploration of this point is the subject of a future article. [31]

At the photon mass-shell point, the coupling constants obtained here at tree-level for both $\gamma\pi\gamma$ and $\gamma\pi\rho$ are within 10% of experiment. Our calculation for $\gamma\pi\rho$ is equivalent to the $0^{th}$ term in the meson loop expansion. An estimate of the meson-loop corrections to our result would be interesting. At this stage we can only remark that, within the same approach to the pion EM form factor, pion-loop corrections were seen to contribute at the level of $< 15\%$ to the charge radius. [32] Similarly, the pion-loop contribution to the $\rho$ mass has the correct sign and magnitude to generate most of the $\rho - \omega$ mass splitting [33] and is a 2% effect. The emerging picture is that a representation of low-mass mesons in terms of dynamically dressed quarks may capture the dominant quantum loop effects. Subsequent meson loop dressing would have to overcome distributed coupling produced by finite size effects.

Instead of considering the underpinnings of the present investigations to be a result of bosonization of the GCM action to the bare meson level, there is a more general viewpoint. [10] Selected truncation of the tower of coupled Dyson-Schwinger equations of QCD, together with use of a generalized impulse approximation, points to the same end result for the processes we consider here. However, the road to higher order effects is different in each. Progression through the loop expansion in the effective action approach defines a rigid ordering of physics content. There is potentially more freedom to develop an efficient ordering with the second approach as one moves beyond the impulse approximation and with less severe DSE truncation. We have taken the former viewpoint to emphasize the immediate relevance to effective hadronic field theory models. It is becoming quite feasible to generate such models with many of the previously phenomenological coupling form factors now given a quark basis that, although approximate, captures the dominant influence on dynamics at the hadron size scale.

**Acknowledgments**    This work has drawn upon a number of valuable previous collaborations with K.L. Mitchell, C.D. Roberts and M.R. Frank. I also thank Amand Fäßler for the organization of a fine program at Erice. This work was supported in part by the National Science Foundation under Grant Nos. INT92-15223 and PHY94-14291.